\documentclass[twocolumn,showpacs,preprintnumbers,amsmath,superscriptaddress,amssymb,pra,aps]{revtex4-1}

\usepackage[utf8]{inputenc}  
\usepackage[T1]{fontenc} 
\usepackage{amssymb}
\usepackage{amsmath}
\usepackage{mathtools}

\begin{document}

\title{Quantitative analysis of losses close to a $d$-wave open-channel Feshbach resonance in Potassium 39}

\author{L. Fouch\'e}
\affiliation{Laboratoire Charles Fabry, Institut d'Optique Graduate School, CNRS, Universit\'e Paris-Saclay, 2 av. A. Fresnel, 91127 Palaiseau Cedex, France}

\author{A. Boiss\'e}
\affiliation{Laboratoire Charles Fabry, Institut d'Optique Graduate School, CNRS, Universit\'e Paris-Saclay, 2 av. A. Fresnel, 91127 Palaiseau Cedex, France}

\author{G. Berthet}
\affiliation{Laboratoire Charles Fabry, Institut d'Optique Graduate School, CNRS, Universit\'e Paris-Saclay, 2 av. A. Fresnel, 91127 Palaiseau Cedex, France}

\author{S. Lepoutre}
\affiliation{Laboratoire Charles Fabry, Institut d'Optique Graduate School, CNRS, Universit\'e Paris-Saclay, 2 av. A. Fresnel, 91127 Palaiseau Cedex, France}

\author{A. Simoni}
\affiliation{Institut de Physique de Rennes, CNRS and Universit\'e de Rennes 1, 35042 Rennes Cedex, France}

\author{T.Bourdel}
\affiliation{Laboratoire Charles Fabry, Institut d'Optique Graduate School, CNRS, Universit\'e Paris-Saclay, 2 av. A. Fresnel, 91127 Palaiseau Cedex, France}



\date{\today}

\begin{abstract}
We study atom losses associated to a previously unreported magnetic
Feshbach resonance in potassium 39. This resonance is peculiar in that it
presents $d$-wave character both in the open and in the closed channels,
directly coupled by the dominant spin-exchange interaction. The losses
associated to a $d$-wave open-channel resonance present specific signatures
such as strong temperature dependance and anisotropic line shapes. The
resonance strength and position depend on the axial projection of the
orbital angular momentum of the system and are extracted from rigorous multichannel
calculations. A two-step model, with an intermediate collision complex
being ejected from the trap after collisions with free atoms, permits to 
reproduce the observed dependance of the loss rate as a function of
temperature and magnetic field.
\end{abstract}

\maketitle

Ultra-cold atoms are many-body quantum systems that offer great control
and versatility \cite{Bloch2008}. Feshbach resonances allow in particular the interatomic interaction to be accurately controlled \cite{Chin2010}.  Such resonances occur when the kinetic energy of two colliding particles in an open channel becomes close to the energy of a bound state in a closed channel potential. Experimentally, Feshbach resonances in atomic collisions are typically induced and controlled using a variable magnetic field, relying on the different magnetic moment of two free atoms and of the resonant molecular state. The main parameter characterizing the interations at ultra-low temperatures (typically below 1$\mu$K), the $s$-wave scattering length $a$, can thus be made to vary and accurately controlled. These
features have permitted the production of weakly bound molecules for
large and positive $a$ \cite{Donley2002, Regal2003a, Cubizolles2003,
Herbig2003, Xu2003, Durr2004}, the study of the BEC-BCS crossover with
fermions \cite{Bartenstein2004, Zwierlein2004, Bourdel2004}, and the
study of resonantly interacting Bose gases \cite{Rem2013, Fletcher2013, Eigen2017}.

In the case of spin-exchange interactions between open and closed
collision channels, the coupling is isotropic and the orbital angular
momentum is conserved. However, other types of coupling such as the
dipolar spin-spin interaction are anisotropic and the orbital momentum can
change. For example, $d$-wave or $g$-wave resonances, where $d$ and $g$
refer to the symmetry of the bound state have been reported for collisions
in the $s$-wave \cite{Marte2002, Chin2002, Chin2010}. Higher partial
wave collisions in the entrance channel can also become resonant at higher energies. These resonances then have
specific features and signatures as the collision rates strongly depends
on the collision energy due to the centrifugal barrier that needs to be
overcome. Feshbach resonances with higher partial waves in the entrance
channel have been reported in $p$-waves \cite{Chin2000, Regal2003b,
Zhang2004} and also in $d$-waves \cite{Beaufils2009, Maier2015, Cui2017}. A $d$-wave
shape resonance was also discovered in $^{41}$K \cite{Yao2017}. Close
to these resonances for fermions, high-order-wave pairing is expected, while $p$-wave
and $d$-wave pairing plays a key role in superfluid liquid $^3$He \cite
{Lee1997} or in $d$-wave Hi-Tc superconductors \cite{Tsuei2000}. For bosons, molecular condensates of rotating molecules are predicted \cite{Yao2018}. Progresses
in these directions have been hindered by the importance of losses and
points toward the need for a quantitative understanding of losses in
the vicinity of the resonances.

In this paper, we report on the observation and quantitative analysis
of a previously unreported Feshbach resonance in potassium 39, that has
$d$-wave character in both the open and the closed channel. We measure and quantitatively model the
associated losses as a function of magnetic field and temperature. The
observed features clearly indicate the $d$-wave nature of the incoming
open channel.

Due to the $d$-wave multiplicity $(l=2, m_l=-2,-1, 0, 1, 2)$, the
resonance is actually composed of five closely-spaced resonances, with
the components with the same $|m_l|$ almost exactly degenerate. The
position, strength and inelastic loss rates associated to each of these two body resonances
are extracted from multichannel calculations based on the collision
model developed in~\cite{DErrico2007}. Our experimental results can
then be $quantitatively$ compared with theoretical predictions from a two step model with first reversible molecule formation and second inelastic losses due to atom-molecule collisions and molecule relaxation. We are able to reproduce both the magnitude and the shape of the loss curves as a function of the magnetic field with a single adjustable parameter , $i.e.$ the collision rate between the quasi-bound resonant molecules and free atoms. In addition, resonant direct three-body processes are found to be unrealistic to explain our data. 

We first prepare a cold gas of $^{39}$K atoms using magneto-optical
trapping and gray molasses working on the D1 atomic transition
\cite{Salomon2014a}. The atoms are then loaded in a strongly
confining crossed optical trap in the $|F=1, m_F=-1\rangle$ state
\cite{Salomon2014b}. Rather than pursuing the subsequent evaporation at
a field of 550\,G to reach condensation, we stop at different trapping
powers in order to prepare thermal gases at different temperatures. The
parameters of the obtained traps and clouds are summarized in table 1
\cite{footnotefreq}.

\begin{table}
\begin{center}
\begin{tabular}{|c|c|c|c|}
  \hline
  $T$ ($\mu$K) & $n_0$ (10$^{19}$m$^{-3}$) & $f_\perp$ (Hz) & $f_\parallel$ (Hz) \\
  \hline
   22 & 5.7 & 3130 & 140   \\
  70 & 3.0 & 5300 & 130   \\
  140 & 1.9 & 7500 & 138  \\
  180 & 1.8 & 8660 & 154  \\
  \hline
\end{tabular}
\end{center}
 \caption{Characteristics of the different samples studied: the temperature $T$, the peak density $n_0$, the frequencies of the dipole crossed trap both in the radial and in the longitudinal directions $f_\perp$ and $f_\parallel$. The uncertainties in temperatures and trapping frequencies are $\sim$10$\%$ . The uncertainties in the densities $\sim 40 \%$ are dominated by the global uncertainty on atom number calibration which is $\sim 30 \%$.}
\end{table}

The magnetic field is ramped down to 408G in 150\,ms and subsequently
precisely tuned between 408\,G and 392\,G in 10\,ms. The atom number
as a function of a variable wait time is then measured by fluorescence
imaging after a few milliseconds time of flight and a sudden switch
off of the magnetic field. The remaining normalized atom numbers after
one second are reported in figure\,1 for the different experimental
conditions. A clear loss feature is observed at 395(1)\,G indicating a
Feshbach resonance. The loss feature is asymmetric. Moreover,
it shifts and broadens with increasing temperature. These features are
experimental evidences that we are dealing with a high-order partial
wave in the entrance channel.

\begin{figure}[htbp]
\centering
\includegraphics[width=0.45\textwidth]{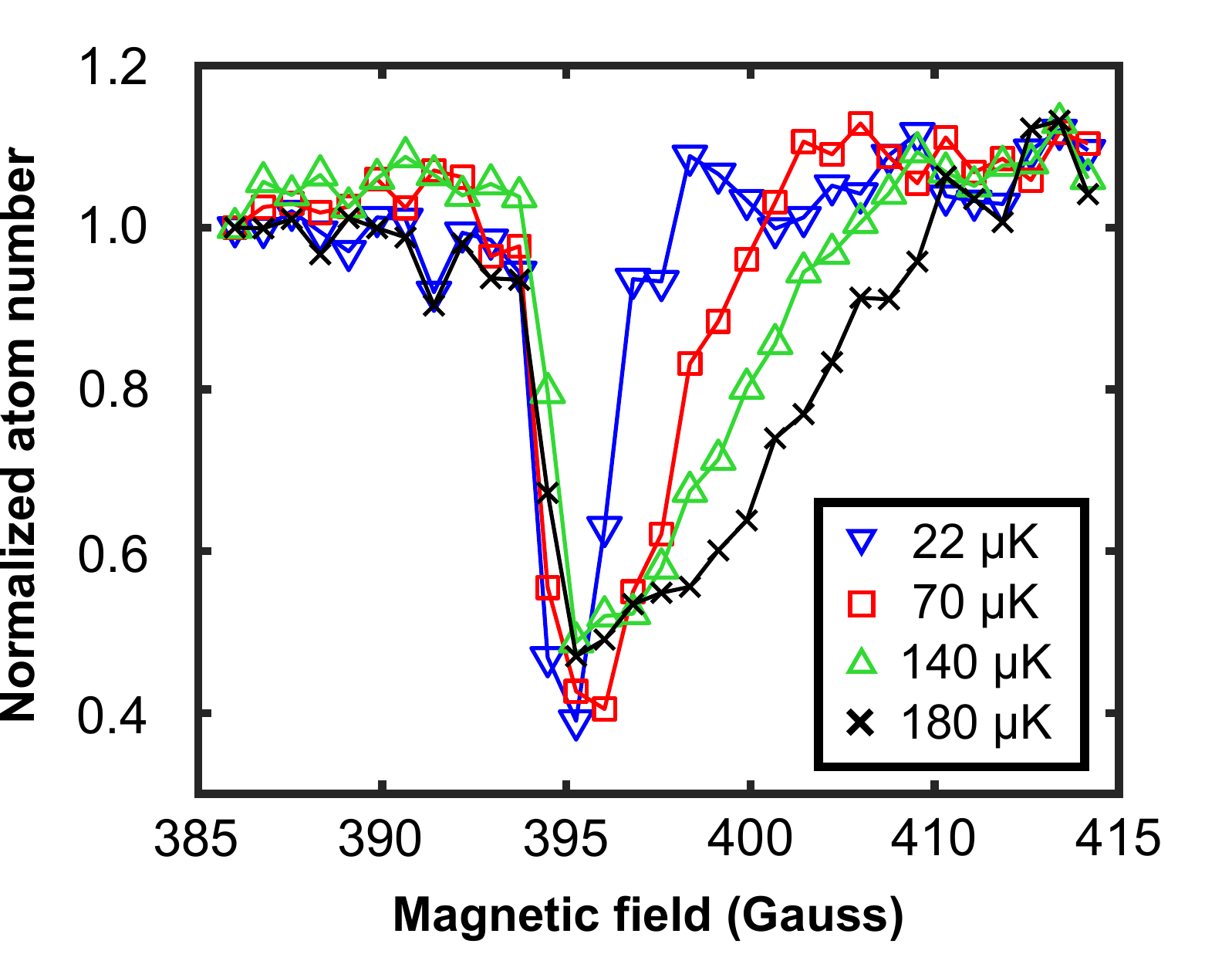}
\caption{Normalized number of atoms remaining in the trap after a waiting time of 1\,s at different magnetic fields. The shape of the loss curve is asymmetric and its width increases with the temperature. (Color online)}
\label{sequence}
\end{figure}

\begin{figure}[htbp]
\centering
\includegraphics[width=0.45\textwidth]{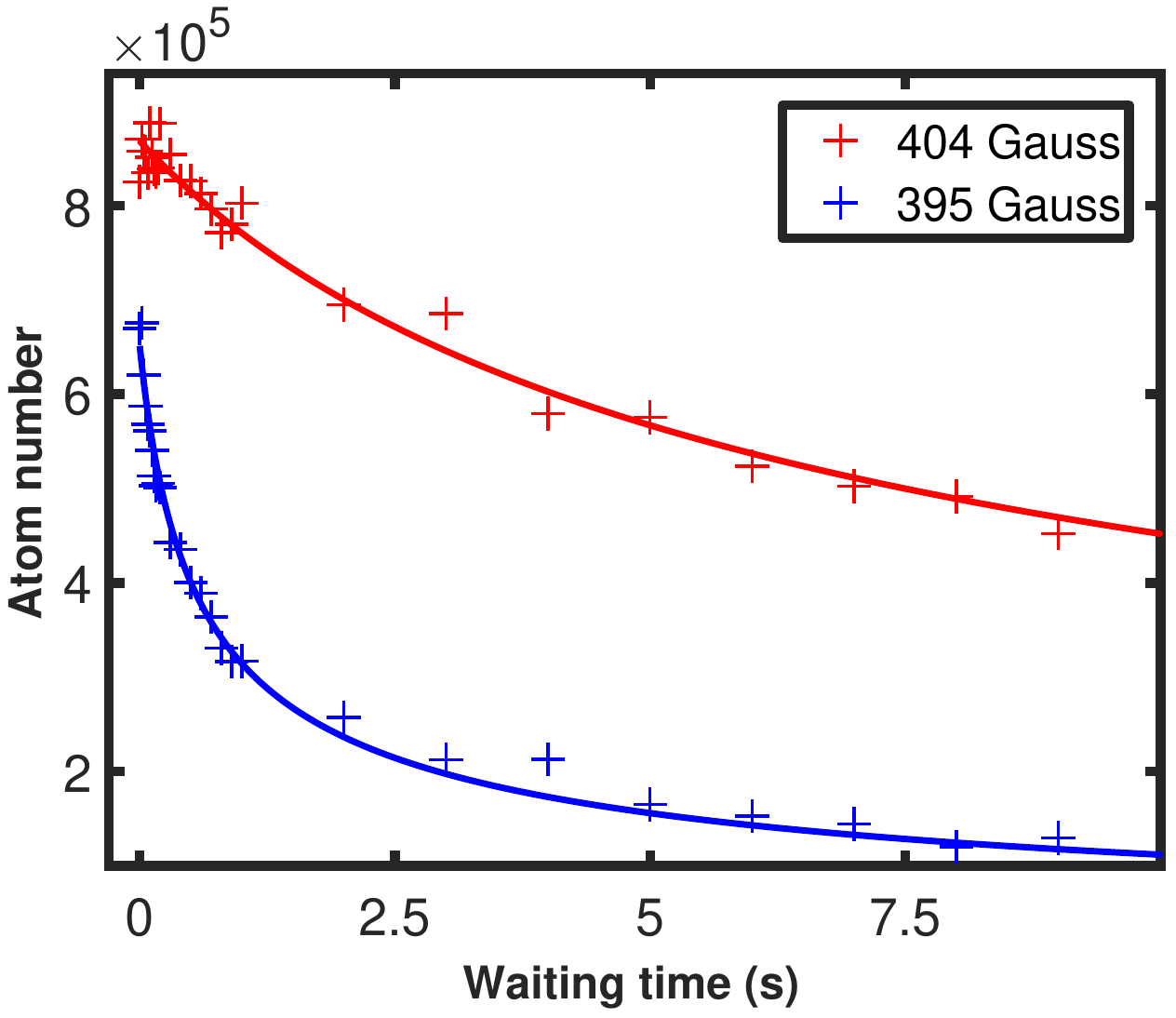}
\caption{Decay of the number of atoms for a trapped gas at 140\,$\mu$K for two different magnetic fields. Experimental points are crosses and fits are solid lines. (Color online)}
\end{figure}
In order to be more quantitative, we now observe the decay curves as
a function of time. In figure\,2, two examples of decay curves at
140\,$\mu$K are plotted. Clearly, the loss rate does increase when approaching the
resonance located at 394\,G. Experimentally, we find that all curves
can be well fitted with a simple three-body plus one-body loss model
\begin{equation}
\frac{1}{N} \frac{dN}{dt}=-\beta_{3} N^{2}-\Gamma_\textrm{1b},
\end{equation}
where $\beta_3$ is the fitted three-body rate constant and $\Gamma_\textrm{1b}=25\,$s$^{-1}$ is the background gas one-body collision loss rate which was measured away from the resonance and that is constant for all data sets. Although the dynamic range and quality of our data is not sufficient to exclude other types of losses such as two-body, the above model gives a good estimate of the initial loss rate, which are the quantity that we latter compare with theory. Assuming a Boltzmann equilibrium density distribution in an harmonic trap, the three-body rate coefficients $K_3$ can be calculated from the values of $\beta_3$ \cite{Weber2003}
\begin{equation}
K_{3}=3^{\frac{3}{2}} \beta_{3}  {\Big(} \frac{k_\textrm{\tiny B} T}{2 \pi m} {\Big)}^3 \frac{1}{f_\perp^4 f_\parallel^2} ,
\end{equation}
where $k_\textrm{\tiny B}$ is the Boltzmann constant and $m$ the atomic
mass.

\begin{figure}[htbp]
\centering
\includegraphics[width=0.45\textwidth]{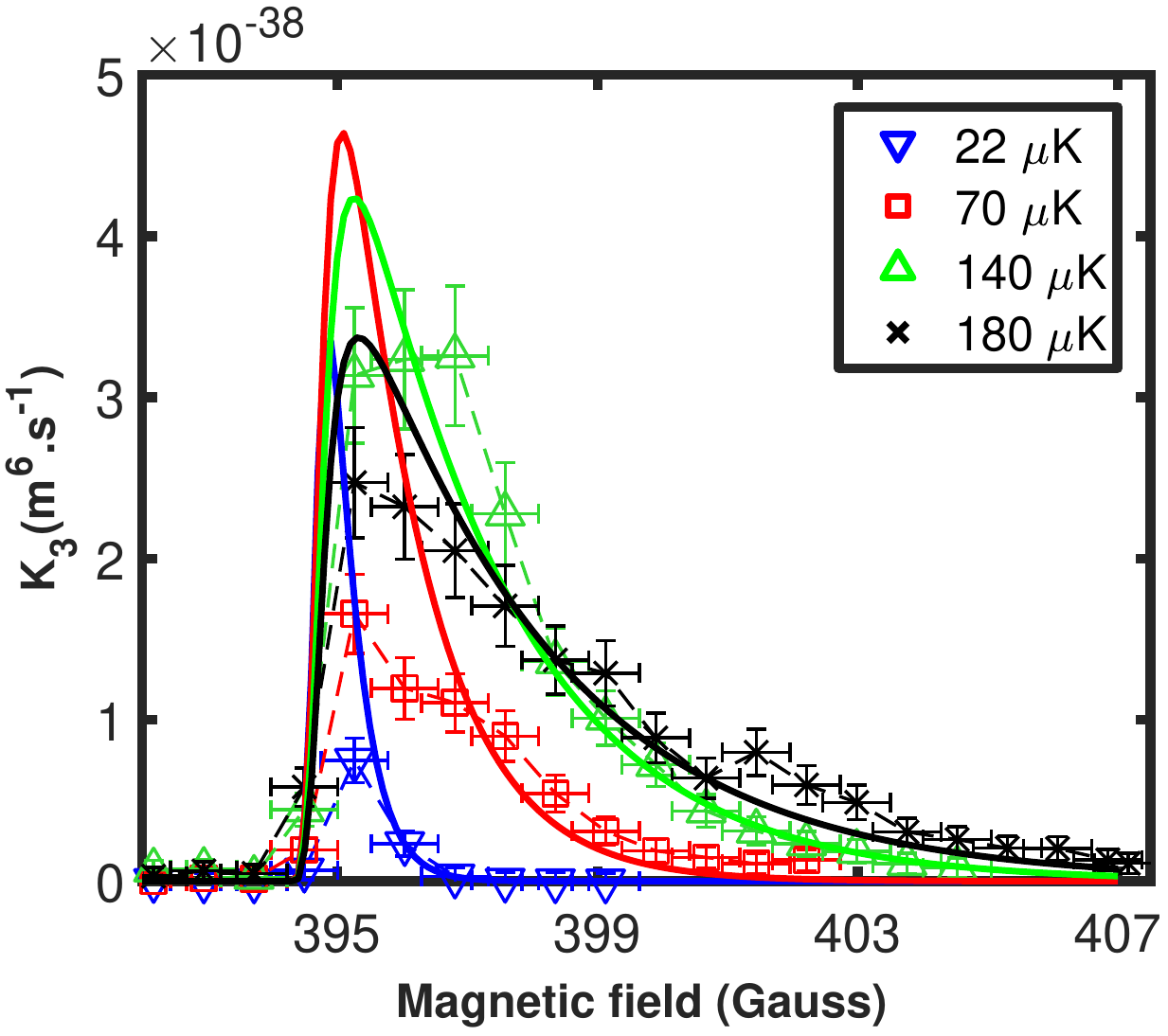}
\caption{
Measured three-body loss rate K$_{3}$ as a function of the magnetic
field. The plotted error bars are the uncertainties estimated from the loss curve fit quality. They do not take into account the uncertainties related to the initial densities which may effect the
absolute values of $K_3$ up to a factor of $\sim$2.5. Ratios of $K_3$
at different temperatures are better controlled with an uncertainty
of $\sim$ 40$\%$. The continuous lines are fits according to the model
described in the text. The fitted values of $C$ are $0, 2.7\times10^{-17}, 6.6 \times 10^{-17},
7.3\times10^{-17}$\,m$^3$.s$^{-1}$ for the sample
with temperatures ranging from 22\,$\mu$K to 180$\,\mu$K. (Color online)}
\end{figure}
The evolution of $K_3$ as a function of the magnetic field is presented
on Fig.\,3 for each sample. The characteristics observed in figure\,1
are retrieved: an asymmetric shape and a width increasing with the
temperature of the sample. Such strong temperature dependance is a consequence of higher partial
wave collisions. In polarized bosonic samples, odd-wave collisions are
forbidden due to the symmetrization principle; it is thus realistic to
believe that our observation is linked to a $d$-wave collision. This
interpretation is confirmed theoretically by the existence of previously unreported $d$-wave
molecular states coinciding in energy at $\sim$394\,G using the scattering
potentials from \cite{DErrico2007}. 


We now turn to a quantitative comparison with the theoretical expectations
for losses associated to the $d$-wave resonances. In the following, we will
use the two-step model described in \cite{Yurovsky2003, Beaufils2009},
each step involving two-body collisions. During the first step, two
atoms collide and are coupled to a molecular quasi-bound state $K_{2} (m)$ in the vicinity
of a Feshbach resonance: 
\begin{equation} 
K + K  \xleftrightarrow {\Gamma ^{el}} K_{2} (m), 
\end{equation} 
with $\hbar \Gamma^{el}$ the energy elastic width. As will be described below in more details, $\Gamma^{el}$ is a quantity linked to two-body physics and it can be accurately calculated. The process is reversible as the resonant molecular state is trapped. Then the molecules may
collide with a third atom, forming a deeply bound dimer $K_{2} (d)$ and releasing a
large amount of energy : 
\begin{equation} 
K_{2} (m) +K \xrightarrow{\Gamma _d} K_{2} (d) + K  
\end{equation} 
In this case, all three atoms involved are lost from the trap.
The energy scale $\hbar \Gamma_{d}$ associated to the molecular lifetime is set by
the inelastic collision with the
surrounding atoms
\begin{equation}
 \hbar \Gamma_{d} (n) = \hbar C n  ,
\end{equation}
where $n$ is the atomic density and $C$ is an atom-molecule collision parameter. Unlike two-body parameters, it is theoretically more difficult
to predict. $C$ is not expected to show a magnetic field dependence as the relaxation process is non-resonant. However $C$ is expected to vary significantly with the atom-molecule collision energy and thus with the temperature in particular due to contributions
from higher order partial waves \cite{Idziaszek2010}. 

In this framework and ignoring non resonant scattering the collisional
cross section $\sigma(k)$ can be represented in the Breit-Wigner
form~\cite{Yurovsky2003} :
\begin{equation} 
\label{eq_BW}
\sigma(k) = \frac{\pi}{k^2} \frac{\hbar^2
\Gamma^{el}(\epsilon)\Gamma_d(n)}{(\epsilon-\epsilon^{\rm res})^2 + \frac{\hbar^2}{4}
\left( \Gamma^{el}(\epsilon) + \Gamma_d(n)  \right)^2} ,
\end{equation} 
where $k$ is the collision relative wave-vector, $\epsilon=\frac{\hbar^2
k^2}{m}$ the collision energy, and $\epsilon^{\rm res}$ the energy location of the Feshbach resonance.

This model was previously used to explain the loss behavior in the
vicinity of the $d$-wave open channel resonance in chromium. In that
case a $l=0$ molecule was coupled to a unique incoming $d$-wave by
the spin dipole interaction, which is relatively strong in chromium \cite{Beaufils2009}.
In our case the situation is more complicated. Our resonant state has
angular momentum $l=2$ and there are thus five participating weakly
bound molecular states that differ by their orbital angular momentum
projection. If we ignore at first the weak spin-spin interaction,
the axial projection of the total hyperfine angular momentum $\vec f =
{\vec F}_a + {\vec F}_b$ on the magnetic field as well as the orbital
angular momentum $\vec l$ are exactly conserved. In this approximation
each metastable state has exactly good quantum labels $\{ m_f , l, m_l\}$
and would decay to a unique spherical wave with the same set of quantum
numbers via the spin-exchange interaction.

The presence of the dipolar interaction slightly complicates the picture,
though still in a perturbative fashion. In fact, the anisotropic character
of the dipolar coupling breaks the separate conservation of $m_f$ and
$\vec \ell$ such that only the axial projection $M$ of the total, orbital
plus hyperfine, angular momentum remains exactly conserved.  However,
since spin-spin mixing with energetically distant states is very weak,
the resonant molecular state retains to excellent approximation $m_f$
and $m_l$ as good quantum number. In general this does not hold in a
weak magnetic field, where states with different projections of $m_f$
are nearly degenerate; see {\it e.g.} Ref.~\cite{Viel2016}. Note that
in our polarized sample the incoming state for the collision as well
as the resonant states have $m_f=-2$. Since at our temperatures only
partial waves up to $l=2$ contribute significantly to the collision,
the relevant total angular momenta range from $M=-4$ to $M=0$. The
spin-spin interaction also introduces two-body losses to lower energy atomic states, which, in addition to three-body processes,
further reduce the lifetime around the resonance.

Position, coupling strengths and inelastic losses can be predicted
very precisely based on a quantum multichannel model comprising
the molecular potentials, hyperfine structure, and dipolar
interaction \cite{DErrico2007}. Calculations have been performed as a
function of total energy $\epsilon$ and magnetic field $B$ in symmetry
blocks labeled by the total axial angular momentum $M$, that will
henceforth be explicitly indicated.

\begin{table*}
\small
\begin{center}
\begin{tabular}{|c|c|c|c|}
  \hline
$M$  & $B^{\rm res}_M$(G)  &$ A_M^{l=2}$ & $\hbar \Gamma_M^{inel}/k_\textrm{B}(\epsilon \to 0)(\mu{\rm K})$     \\
 \hline
-4   &394.35  & 0.20 & 0   \\
-3   &394.50  & 0.20 & $1.51\times 10^{-3}$   \\
-2   &394.54  & 0.20 & $8.58\times 10^{-3}$  \\
-1   &394.50  & 0.20 & $1.33\times 10^{-2}$   \\
 0   &394.35  & 0.20 & $1.05\times 10^{-2}$   \\
  \hline
\end{tabular}
\end{center}
\label{tab_params}
\caption{Resonance parameters of the $d$-wave resonance multiplet
resolved for total angular momentum $M$. The zero-energy magnetic field
location $B^{\rm res}_M$ of the resonance, the coefficient $A_M^{l}$ of
the threshold expansion of the $\ell=2$ elastic width, and the inelastic
width $\Gamma_M^{inel}$ computed at the threshold energy of the incoming
atoms are shown (see text).}
\end{table*}

\begin{figure}[htbp]
\centering
\includegraphics[width=0.45\textwidth]{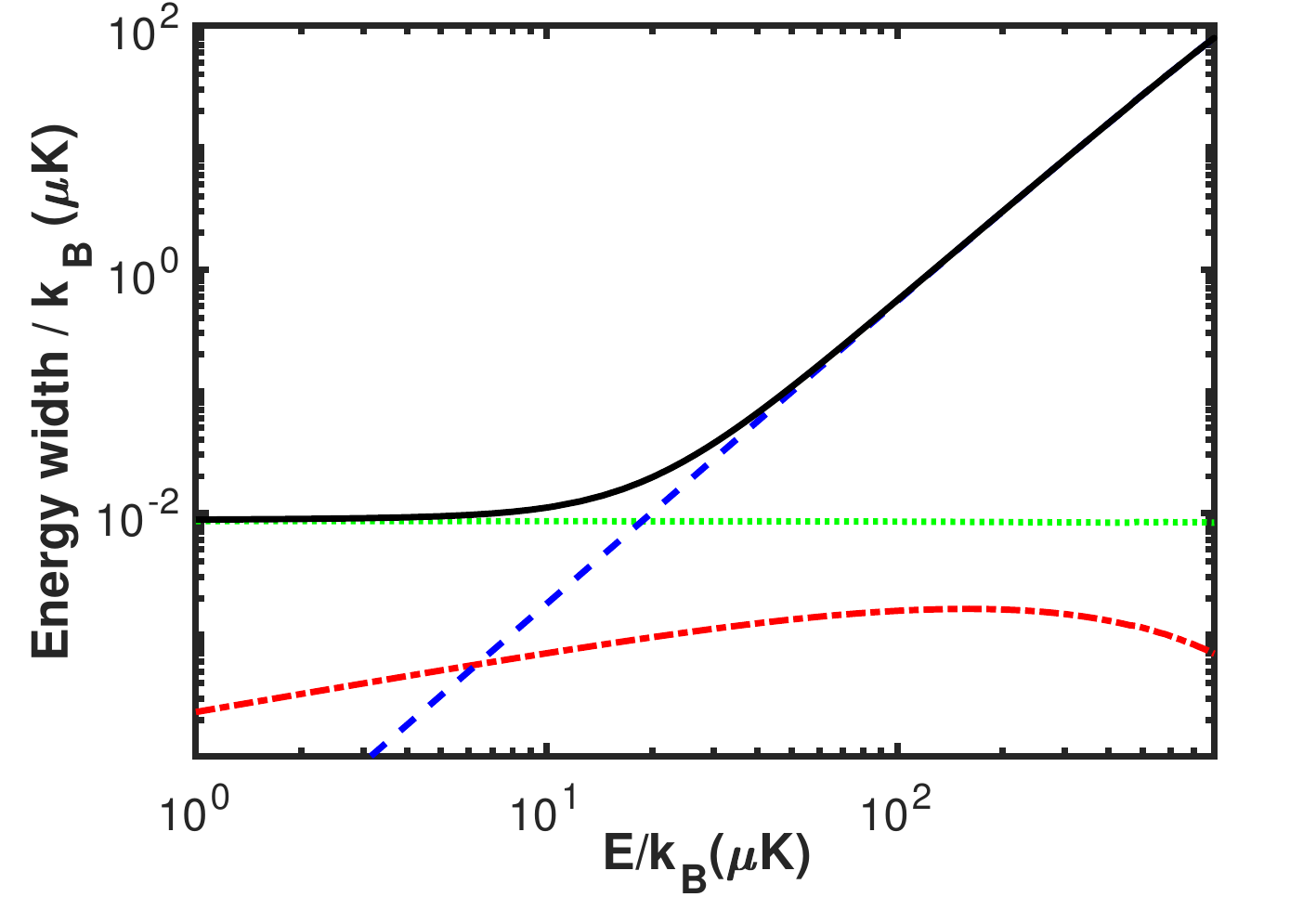}
\caption{Partial decay widths of a $l=2$ resonance to the $\{ \alpha_{in}
,l=2\}$ (dashed line), $\{ \alpha_{in}, l=0\}$ (dotted line)
elastic collision channels as a function of the resonance position. The
total inelastic width (dash-dotted line) and the full width (full
line) are also shown (see text for details). Calculation is for total
magnetic quantum number $M=-2$.} \label{fig_gamma} \end{figure} In
order to extract from the numerical calculation the partial widths for
dissociation into different channels we adopt the time delay operator,
defined in terms of the full scattering matrix $\bm S$ as~\cite{Smith1960}
\begin{equation} 
{\bm Q}_M(\epsilon) = i \hbar {\bm S}_M \frac{{\mathrm
d}{\bm S}^\dagger_M}{{\mathrm d}\epsilon} .
\end{equation}
The computational details are too long to convey here, we simply
mention that the energy derivatives of the $\bm S$ matrix are obtained
analytically in the framework of the spectral element representation of
the Hamiltonian~\cite{Simoni2017}.

The diagonal terms of ${\bm Q}_M$ correspond to the average time delay
experienced during a collision starting in a given channel, whereas
its eigenvalues $q_i$ are associated with the lifetime of metastable
states of the molecular system. Large eigenlifetimes, namely such
that $q_i \gg \hbar / \epsilon $ correspond to long-lived resonant
states~\cite{Smith1960}. In particular, near an isolated resonance
occurring away from energy thresholds the maximum eigenvalue of the
lifetime operator with eigenvector ${\bm v}_M(\epsilon)$ exhibits a
Lorentzian dependence on collision energy
\begin{equation}
\label{eq_td}
q^{\rm max}_{M}(\epsilon)= \frac{\hbar^2 \gamma_{M}}{(\epsilon-\epsilon_M^{\rm res})^2 +\hbar^2 \gamma_{M}^2/4} ,
\end{equation}
with $\hbar \gamma_{M} = \hbar /  q^{\rm max}_M(\epsilon_M^{\rm res})$ the total
energy width. 

As discussed in Ref.~\cite{Simoni2009}, when the resonance occurs at
low energy the collision lifetime will in general be distorted from
the simple Lorentzian profile due to energy threshold effects embedded
in $\gamma$ and $\epsilon^{\rm res}$. In the present case however the
considered resonances are sufficiently narrow that the energy variation
of such parameters over the resonance width can be ignored and Eq.~(\ref{eq_td}) remains
accurate. Moreover, it can be shown that the decay probability into a
specific channel $\{ \alpha, l,m_l \}$ is given by the squared eigenvector
component $P^{\alpha l m_l}_M= |v^{\alpha l m_l}_M(\epsilon_M^{\rm
res})|^2$~\cite{Smith1960}. The partial width on resonance is then
$\gamma^{\alpha l m_l}_M = P^{\alpha l m_l}_M \gamma_{M}$. For notational
convenience we will let $\alpha_{in}$ the internal quantum numbers of
the colliding atoms, namely, in the case of our polarized sample, $
\alpha_{in}  = \{ F_a=1, m_{Fa}=-1 , F_b=1, m_{Fb}=-1 \}$.

At the temperatures of the current experiment, the energy dependence of
the partial width essentially depends on the centrifugal barrier in the
exit channel. For a channel at threshold with angular momentum $l$
this behavior amounts to the Wigner law $\gamma^{\alpha l m_l}_M \sim
\epsilon^{l +1/2 }$. We find it convenient to express such scaling law
in the form $\hbar \gamma^{\alpha l m_l}_M = A^l_M \epsilon_{\rm
vdW} (\epsilon / \epsilon_{\rm
vdW})^{l +1/2 }$, with $A^l_M$ a dimensionless coefficient, $\epsilon_{\rm
vdW} = k_\textrm{B}\times 1.06~{\rm mK}$ the characteristic energy of the van der Waals
potential for potassium~\cite{Chin2010}. Decay to deeper inelastic
channels is weakly dependent on collision energy on the $\mu$K scale.
Note that decay to the $s$-wave is only possible for $M=-2$, the other $M$
projections are only coupled to elastic $d$-wave channels or to inelastic
channels. The $M=-4$ resonant component represents an exception in that
it can only decay to $\{ \alpha_{in} , l=2, m_l=-2\}$ since no competing
inelastic channels exist for this $M$ value.

Based on the partial widths, we define a total two-body elastic width
to access the resonance state from channel $\alpha_{in}$
\begin{equation}
\hbar \Gamma_M^{el}=\hbar \sum_{\ell m_l} \gamma_M^{\alpha_{in} \ell m_l}
\end{equation}
and an inelastic one to leave the resonance by two-body decay
towards all energetically open channels but the incoming one :
\begin{equation}
\hbar \Gamma_M^{inel}=\hbar \sum_{\alpha \neq \alpha_{in}} \sum_{\ell m_l} \gamma_M^{\alpha \ell m_l} .
\end{equation}

The threshold behavior is confirmed by inspection of Fig.~\ref{fig_gamma},
that depicts the elastic and inelastic partial widths for the sample
value $M=-2$. Note that the total width is mostly controlled by the
elastic $\{ \alpha_{in} , l=2, m_l \}$ channel for energies above
few tenths of $\mu$K. We have checked that the dipolar interaction has a
negligible influence on the decay rate for such elastic decay pathway thus
confirming that it proceeds by spin-exchange  \cite{comparisonGao}. On the converse, coupling
to the $s$-wave $\{ \alpha_{in} , l=0, m_l=0 \}$ solely results from the
weak spin-spin interaction but also gives a minor contribution to the total
resonance width. Inelastic decay is dominant at small energies
$\lesssim 10\mu$K but overall is a slow process since our initial state is
stable under spin-exchange and only decays through weak magnetic dipolar
interactions. The inelastic width is otherwise weakly dependent on energy.

Quantitative resonance parameters extracted from the time-delay formalism
for each $M$ can be found in table~2. The table also
contains the zero-energy magnetic field location of the multiplet
components, related to the resonance energy by $\epsilon_M^{\rm
res}=\delta \mu (B - B_M^{\rm res})$, with $\delta \mu$ the relative
magnetic moment of the resonant state with respect to the separated atoms.
The latter depends on the internal spin structure of the metastable
molecule and its magnitude $\delta \mu =60~\mu$K/G is to very good
approximation independent of the particular multiplet component
considered. Note that the $\ell =2$ partial width coefficient $A_M^\ell$
is also essentially independent of $M$, as it can be expected 
according to the approximate conservation of $\vec l$ for weak anisotropic interactions.

Using the previous considerations, introducing the sum over independent resonance components we can define two Breit-Wigner cross sections
\begin{gather}
\sigma_d(k) =
\sum_M \frac{\pi \hbar^2
\Gamma^{el}_M(\epsilon)\Gamma_d(n)/k^2}{(\epsilon-\epsilon^{\rm res}_M)^2 + \frac{\hbar^2}{4}  \left( \Gamma^{el}_M(\epsilon) + \Gamma_d(n) +\Gamma_M^{inel}  \right)^2}
\\
\sigma_{inel}(k) =
\sum_M \frac{\pi \hbar^2
\Gamma^{el}_M(\epsilon)\Gamma_M^{inel}/k^2}{(\epsilon-\epsilon^{\rm res}_M)^2 + \frac{\hbar^2}{4}  \left( \Gamma^{el}_M(\epsilon) + \Gamma_d(n) +\Gamma_M^{inel}  \right)^2}, 
\end{gather}
where $\sigma_d(k)$ corresponds to losses induces by atom-molecule collision and $\sigma_{inel}(k)$ corresponds to losses induced by molecule inelastic relaxation. 
Here, we have assumed for simplicity that $\Gamma_d(n)$ is independent of $M$ as we do not expect very different values of the relaxation toward deeply bound states. The atom loss rate at temperature $T$ can then be calculated by 
averaging over the Maxwell-Boltzmann distribution of atoms. 
Similarly to the
work on chromium~\cite{Beaufils2009} we assume that the resonance widths
$\hbar(\Gamma_M^{el}+\Gamma_d+\Gamma_M^{inel})$ are much smaller than the temperatures such
that the denominator of the cross-sections can be replaced by a Dirac $\delta$-function. However,
in contrast to the approximation done in the chromium paper, in order
to properly account for our data we do not assume any relation between
$\Gamma^{el}, \Gamma_d$ and $\Gamma_M^{inel}$. The resulting initial loss rate can be formulated
as an effective three body coefficient $K_3^\textrm{th}$ which depends on the local density $n$ (and thus also incorporate two-body losses)
\begin{equation}
K_3^\textrm{th}=\Big(\frac{4\pi\hbar^2/m}{k_b T}\Big)^{\frac{3}{2}}\sum_{M}  \frac{\Gamma_{M}^{ el} (3C+2\Gamma_{M}^{ inel}/n)}{\Gamma_{M}^{el}+\Gamma_{M}^{inel}+C n} \textrm{e}^{-\epsilon^{\rm res}_{ M}/k_b T},
\end{equation}
where the factor 2 and 3 originates from the number of lost atoms in each processes. In the regime $\Gamma^{el} \gg C n \gg \Gamma^{inel}$, $K_3^\textrm{th}$ is then independent of density and correspond to a three-body loss behavior. In order to compare with the experimental finding, we calculate the initial loss rate by integration over the trap volume. The comparison as a fonction of magnetic field for the different temperature data sets is rather presented in terms of $K_3$ calculated as if there were only three-body losses as in the experimental fits (see Fig.\,3). 

The only unknown parameters are $C$ values, which we fit to our data sets at each temperature independently. We are
able to reproduce to a relatively good accuracy the experimental loss curves as shown in figure 3. In particular, both the overall shape of each curve as well as the scaling between different temperature data sets are respected. The fact that experimental data close to resonance are always below theoretical expectations can be understood: For these points, the initial loss rates are larger than the longitudinal trap frequencies and the high density region of the trap are quickly depleted reducing the losses. The hypothesis of thermal equilibrium is no longer valid.

The values of the fitted $C$ parameters are given in the figure
caption. The results are sensitive to the calibration of the atom number and thus to a global scaling of the densities. If the densities are underestimated by 40$\%$, thus giving smaller values of $K_3$ the fitted value $C$ tend to zero. This is not really realistic as in this case, the scalings between the different temperature data sets are not as well reproduced. On the contrary, if the densities are overestimated by 40$\%$, leading to higher values of $K_3$, the fits are as good with values of $C$ that are higher by a factor up to 4. At 22\,$\mu$K, the data are best fitted by pure two-body inelastic losses (i.e. $C=0$) but a value $C$ of the order of $10^{-17}$\,m$^3$.s$^{-1}$ is also within experimental uncertainties. The fitted values of $C$ can be compared to the universal rates described
in \cite{Idziaszek2010} assuming full reactivity at short range. Taking
an atom-molecule $C_6$ coefficient equal to twice the atom-atom $C_6$,
we calculate a universal value for $s$-wave reactions $\sim$1.2$\times
10^{-16}\,$m$^3$ s$^{-1}$ which is indeed of a similar scale as
the experimental finding. Moreover, the observed increasing values of $C$ as a
function of temperature a result that is robust against a global error on the atom number calibration can be expected from $p$-wave atom-molecule
collisions that have been shown to contribute significantly at our
energies~\cite{Idziaszek2010}. 

The above loss model is able to accurately reproduce our experimental finding with realistic values of the atom-molecule inelastic collision rates which are the only parameters. Nevertheless, one may wonder about the possibility to fit our data with direct three-body relaxation, which is another decay mechanism close to a $d$-wave resonance \cite{Wang2012}, for example put forward to explain losses close to a $d$-wave resonance in Erbium \cite{Maier2015}. At 180\,muK, the unitary three-body loss coefficient \cite{Fletcher2013} is 2.4$\times 10^{-38}$m$^6.$s$^{-1}$ (including a factor 5 due to the resonance multiplicity), a value comparable to our experimental measurement. However, the abrupt increase of losses as a function of magnetic field, together with the  losses increasing with temperature are not consistent with a unitary limited regime. We indeed find that it is not possible to reproduce our data (both in magnitude and shape) with resonant three-body processes as described in \cite{Maier2015}. Such processes are thus probably not significantly contributing to our observed losses.


In conclusion, we have studied a $d$-wave resonance at 394 G in potassium
39 in the $|1, -1\rangle$ hyperfine state. The dependance of atomic
losses as a function of temperature and magnetic field proves that the
molecular states are coupled to a $d$-wave open channel. More precisely,
the five components of the incoming $d$-wave channel are isotropically
coupled through spin exchange to five different molecular states with
$d$-wave symmetry and very close in energy. The coupling to these
states have been calculated theoretically. With realistic values of the
collision rate between molecular states and free atoms as a function
of energy, we can reproduce the values of the loss rates as a function
of magnetic field and temperature. Our results permit to precise the type of loss processes close to $d$-wave Feshbach resonances and could lead to interesting developments toward the
observation of $d$-wave pairing for fermions and of
strongly interacting $d$-wave superfluid Bose gases \cite{Yao2018}. Our results
indicate the need to work at a low density in order to avoid molecule-atom
relaxation.

\acknowledgments
This research has been supported by CNRS, Minist\`ere de l'Enseignement Sup\'erieur et de la Recherche, Direction G\'en\'erale de l'Armement, ANR-12-BS04-0022-01 and ANR-12-BS04-0020-01, Labex PALM, ERC senior grant Quantatop, Region Ile-de-France in the framework of DIM Nano-K (IFRAF), EU - H2020 research and innovation program (Grant No. 641122 - QUIC), Triangle de la physique.


\begin{thebibliography}{0}
\bibitem{footnotefreq}
The longitudinal frequencies are effective frequencies estimated
from thermal density profiles in the potential which originates from
both optical trapping beams.
\end{thebibliography}
\end{document}